\documentstyle[12pt,epsf]{article}
\newcommand{\Z}{{\sf Z \!\!\! Z}}
\newcommand{\1}{{\sf 1 \!\! 1}}
\newcommand{\p}{\partial}
\makeatletter
\@addtoreset{equation}{section}
\makeatother

\title{Can One See the Number of Colors ?
\footnote{This work is supported in part by funds provided by the U.S.
Department of Energy (D.O.E.) under cooperative research agreement
DE-FC02-94ER40818.}}

\author{O. B\"ar and U.-J. Wiese\footnote{e-mail: 
obaer@mitlns.mit.edu, wiese@mitlns.mit.edu} \\ \\
Center for Theoretical Physics, \\
Laboratory for Nuclear Science and Department of Physics \\
Massachusetts Institute of Technology (MIT) \\
Cambridge, Massachusetts 02139, U.S.A.}

\begin{document} 
\maketitle

\begin{center}

MIT preprint: MIT-CTP 3144


\end{center}

\begin{abstract} \normalsize

We formulate the standard model with an arbitrary number of colors $N_c$. The 
cancellation of Witten's global $SU(2)_L$ anomaly requires $N_c$ to be odd,
while the cancellation of triangle anomalies determines the consistent 
$N_c$-dependent values of the quark charges. In this theory, the width of the 
decay $\pi^0 \rightarrow \gamma \gamma$ is not proportional to  $N_c^2$. In 
fact, in the case of a single generation and hence for two quark flavors
($N_f = 2$), $N_c$ does not appear explicitly in the low-energy effective 
theory of the standard model. Hence, contrary to common lore, it is impossible 
to see the number of colors in low-energy experiments with just pions and 
photons. For $N_f \geq 3$, on the other hand, $N_c$ explicitly enters the 
chiral Lagrangian as the quantized prefactor of the Wess-Zumino-Witten term,
but the contribution of this term to photon-pion vertices is completely 
canceled by the $N_c$-dependent part of a Goldstone-Wilczek term. However, the
width of the decay $\eta \rightarrow \pi^+ \pi^- \gamma$ survives the
cancellation and is indeed proportional to $N_c^2$. By detecting the emerging
photon, this process thus allows one to literally see $N_c$ for $N_f \geq 3$.

\end{abstract}
 
\maketitle
 
\newpage

\section{Introduction}

In this article we ask if it is possible to deduce the number of colors 
directly from low-energy experiments with photons, pions, $\eta$-mesons and 
kaons. For example, can one literally see $N_c$ by detecting the photons 
emerging from the decay of a neutral pion? Unfortunately, there are several 
misleading statements in the literature concerning this question. First, one 
often reads that the standard model is anomalous --- and hence inconsistent ---
for $N_c \neq 3$. Then one reads in almost any textbook on the subject that the
width of the decay $\pi^0 \rightarrow \gamma \gamma$ is proportional to 
$N_c^2$, and that the observed width is consistent only with $N_c = 3$. It was 
pointed out by Abbas that both of these statements are wrong 
\cite{Abb90,Abb00}, simply because varying $N_c$ without adjusting the quark 
charges accordingly is inconsistent. Actually, these first two standard pieces 
of evidence for $N_c = 3$ do not at all imply that there are three colors. The 
third standard textbook evidence for three colors is provided by the Drell 
ratio $R$. When evaluated at high energies, this ratio is sensitive to $N_c$ 
and its measured value indeed implies $N_c = 3$. Here we ask if there are also 
low-energy processes which allow one to directly see $N_c$, and which could 
hence replace the misleading textbook example 
$\pi^0 \rightarrow \gamma \gamma$.

In order to cancel Witten's global anomaly \cite{Wit82}, the number of colors 
must be odd in the standard model. The cancellation of triangle anomalies 
requires the electric charges of the up and down quarks to be \cite{Abb90}
\begin{equation}
\label{charges}
Q_u = \frac{1}{2} (\frac{1}{N_c} + 1), \quad 
Q_d = \frac{1}{2} (\frac{1}{N_c} - 1).
\end{equation}
For $N_c = 3$ these are the familiar values $Q_u = \frac{2}{3}$ and $Q_d = 
- \frac{1}{3}$. Keeping these charges fixed while varying $N_c$ is 
inconsistent, because the anomalies no longer cancel. Of course, in a 
vector-like theory with only electromagnetic and strong gauge interactions, any
electric charge assignment is consistent with any choice of $N_c$. In a chiral 
gauge theory with electroweak gauge interactions, on the other hand, anomaly 
cancellation leads to eq.(\ref{charges}). If one already knows the quark 
charges to be $Q_u = \frac{2}{3}$ and $Q_d = - \frac{1}{3}$ (which is what the 
textbooks implicitly do), both anomaly cancellation and the $\pi^0$ decay 
indeed imply $N_c = 3$. However, in that case, one could simply say that the 
observed charges of the proton and the neutron already imply three colors.

In a world with $N_c = 5$ colors, the quark charges are $Q_u = \frac{3}{5}$ and
$Q_d = - \frac{2}{5}$. In such a world, baryons consist of five quarks. For 
example, the proton now contains three up quarks and two down quarks, but still
has electric charge $Q_p = 3 Q_u + 2 Q_d = 1$. The neutron consists of two up 
quarks and three down quarks and still is electrically neutral. For arbitrary 
odd $N_c$, the proton is made of $(N_c + 1)/2$ up quarks and $(N_c - 1)/2$ down
quarks, while the neutron contains $(N_c - 1)/2$ up quarks and $(N_c + 1)/2$ 
down quarks. Hence, as in our world,
\begin{eqnarray}
&&Q_p = \frac{N_c + 1}{2} Q_u + \frac{N_c - 1}{2} Q_d = 1, \nonumber \\
&&Q_n = \frac{N_c - 1}{2} Q_u + \frac{N_c + 1}{2} Q_d = 0.
\end{eqnarray}
There is also still a $\Delta$-isobar with the usual electric charges
\begin{eqnarray}
&&Q_{\Delta^{++}} = \frac{N_c + 3}{2} Q_u + \frac{N_c - 3}{2} Q_d = 2, 
\nonumber \\ 
&&Q_{\Delta^+} = \frac{N_c + 1}{2} Q_u + \frac{N_c - 1}{2} Q_d = 1, 
\nonumber \\
&&Q_{\Delta^0} = \frac{N_c - 1}{2} Q_u + \frac{N_c + 1}{2} Q_d = 0, 
\nonumber \\
&&Q_{\Delta^-} = \frac{N_c - 3}{2} Q_u + \frac{N_c + 3}{2} Q_d = - 1.
\end{eqnarray}
In general, in a world with two flavors and an arbitrary odd number of colors,
baryons have equal half-integer valued isospin and spin $1/2 \leq I = S  \leq
N_c/2$. For example, for $N_c = 5$ there is an additional baryon resonance 
beyond the $\Delta$-isobar with $I = S = 5/2$. For arbitrary odd $N_c$ the 
highest of these additional resonances has $I = S = N_c/2$. The member of this 
multiplet with the largest electric charge consists of $N_c$ up quarks and has 
$Q = (N_c + 1)/2$ while the member with the most negative charge contains 
$N_c$ down quarks and has $Q = - (N_c - 1)/2$. Obviously, the high end of the 
baryon spectrum is sensitive to the number of colors. On the other hand, the 
familiar baryons $p$, $n$ and $\Delta^{++}$, $\Delta^+$, $\Delta^0$, $\Delta^-$
with their usual electric charges exist for any odd $N_c \geq 3$. 

Of course, the interior of a baryon with $N_c \neq 3$ is different from that of
a baryon in our world. This definitely has observable consequences at short 
distances and thus at high energies. For example, the up, down and strange 
quark contribution to the Drell ratio (with $Q_s = Q_d$)  is proportional to
\begin{equation}
R = \frac{\sigma(e^+ e^- \rightarrow \mbox{hadrons})}
{\sigma(e^+ e^- \rightarrow \mu^+ \mu^-)} \propto
N_c (Q_u^2 + 2 Q_d^2) = \frac{3}{4}(N_c + \frac{1}{N_c} - \frac{2}{3}).
\end{equation}
Instead, if one uses $Q_u = \frac{2}{3}$ and $Q_d = - \frac{1}{3}$ independent 
of $N_c$, one obtains the inconsistent textbook result $R \propto \frac{2}{3} 
N_c$. Still, in either case, $R$ is quite sensitive to the number of colors. 
However, the above calculation of the Drell ratio is valid only at high 
energies, above about 1 GeV. Here we ask if and how $N_c$ enters the low-energy
electromagnetic physics of pions, $\eta$-mesons and kaons. We will find that, 
using the consistent charge assignment for the quarks of eq.(\ref{charges}), 
several anomalous processes are $N_c$-independent. This contradicts common 
lore, which says that, for example, the decay width of 
$\pi^0 \rightarrow \gamma \gamma$ is proportional to $N_c^2$, thus changing 
drastically with the number of colors. 

The decay $\pi^0 \rightarrow \gamma \gamma$ results from a triangle diagram 
with an internal quark loop attached to two external electromagnetic 
$U(1)_{em}$ currents and one external isovector axial current. This diagram is 
proportional to
\begin{equation}
\label{trace}
\mbox{Tr}(T^3 Q^2) = \frac{N_c}{2} (Q_u^2 - Q_d^2),
\end{equation}
where $T^3 = \frac{1}{2} \tau^3$ is the diagonal generator of isospin, and 
$Q = \mbox{diag}(Q_u,Q_d)$ is the quark charge matrix. Note that the trace in
eq.(\ref{trace}) implies a sum over color indices. As we will see later, the 
triangle diagram gives rise to the effective vertex
\begin{equation}
{\cal L}_{\pi^0 \gamma \gamma} = - i N_c (Q_u^2 - Q_d^2) 
\frac{e^2}{32 \pi^2 F_\pi}
\pi^0 \varepsilon_{\mu\nu\rho\sigma} F_{\mu\nu} F_{\rho\sigma},
\end{equation}
where
\begin{equation}
F_{\mu\nu} = \p_\mu A_\nu - \p_\nu A_\mu
\end{equation}
is the field strength of the electromagnetic vector potential $A_\mu$. Using 
$Q_u = \frac{2}{3}$ and $Q_d = - \frac{1}{3}$ this leads to the textbook result
for the decay width
\begin{equation}
\Gamma(\pi^0 \rightarrow \gamma \gamma) = 
\left(\frac{N_c}{3}\right)^2 \frac{e^4 M_\pi^3}{1024 \pi^5 F_\pi^2},
\end{equation}
which is proportional to $N_c^2$. This result compared to the observed width 
seems to imply $N_c = 3$. However, by fixing the quark charges to $Q_u = 
\frac{2}{3}$ and $Q_d = - \frac{1}{3}$, one has already implicitly assumed that
$N_c = 3$. On the other hand, if one uses the consistent quark charges given in
eq.(\ref{charges}), one obtains
\begin{equation}
N_c (Q_u^2 - Q_d^2) = \frac{N_c}{4} [(\frac{1}{N_c} + 1)^2 - 
(\frac{1}{N_c} - 1)^2] = 1,
\end{equation}
and the width turns into
\begin{equation}
\label{width}
\Gamma(\pi^0 \rightarrow \gamma \gamma) = 
\frac{e^4 M_\pi^3}{1024 \pi^5 F_\pi^2},
\end{equation}
independent of $N_c$. Of course, the pion mass $M_\pi$ and the pion decay
constant $F_\pi$ also depend on $N_c$ implicitly. However, if one takes $M_\pi$
and $F_\pi$ from experiment, the implicit dependence is irrelevant and the 
observed width does not imply $N_c = 3$. On the other hand, if one computes 
$M_\pi$ and $F_\pi$ using lattice QCD for different values of $N_c$ --- which 
is certainly a nontrivial task --- one would still infer $N_c = 3$. However,
using that method, one can deduce the number of colors from any QCD observable,
not only from processes like the $\pi^0$ decay.

Since the 1970s, the misleading statement that the $\pi^0 \rightarrow \gamma 
\gamma$ decay width is proportional to $N_c^2$ has been used to lend support to
the correct conclusion that in our world $N_c = 3$. In this context it is 
sometimes stated that Steinberger, who obtained almost the correct width from a
nucleon triangle diagram as early as 1949 \cite{Ste49}, accidentally got the 
right answer from an incorrect theory. From the point of view presented in this
paper, Steinberger's result is the correct answer that one obtains from any 
low-energy effective theory of the standard model even if $N_c \neq 3$. His
result is what one obtains for $N_c = 1$, i.e.\ in a colorless world without 
quarks. Of course, in 1949 Steinberger did not know about quarks or color, but 
he was still using a consistent low-energy effective description of our world. 
Indeed, the standard model with $N_c = 1$, and hence without strong 
interactions, is also anomaly free. According to eq.(\ref{charges}) the 
``quark'' charges are then equal to $Q_u = 1$ and $Q_d = 0$, and the up and 
down ``quarks'' are, in fact, just the proton and the neutron. A standard model
with the usual Higgs and electroweak sector, but with $N_c = 1$ has a nucleon, 
but no strong interactions, no spontaneous chiral symmetry breaking, and thus 
no pions. Still, one can add the pions in the form of a Gell-Mann Levy linear 
$\sigma$-model \cite{Gel60} without spoiling renormalizability or anomaly 
cancellation. In such a model, which is close to what Steinberger used, one 
again obtains eq.(\ref{width}) despite the fact that there are no colors at 
all.

The low-energy physics of the strong interactions is governed by the 
pseudo-Goldstone bosons of spontaneous chiral symmetry breaking --- the pions,
$\eta$-mesons and kaons. Chiral perturbation theory offers a systematic 
approach to describe the Goldstone boson dynamics at low energies \cite{Gas85}.
In QCD with $N_f$ massless quark flavors, the chiral symmetry is 
$SU(N_f)_L \otimes SU(N_f)_R$ which breaks spontaneously to $SU(N_f)_{L=R}$. 
This should be the case for any number of colors, except for $N_c = 2$. The 
$N_c = 2$ case is special because then quarks and anti-quarks are 
indistinguishable and the chiral symmetry is $SU(2 N_f)$, which is expected to 
break spontaneously to $Sp(2 N_f)$ \cite{Pes80}. For any even $N_c$ the baryons
are bosons and hence their physics is qualitatively different from that of the 
real world. Here we are interested in generalisations of the standard model to 
$N_c \neq 3$ that are at least qualitatively similar to our world. 
Interestingly, as mentioned before, the cancellation of Witten's global 
anomaly, separately for each generation, already limits us to odd $N_c$. In 
that case, the Goldstone bosons are described by fields in the coset space 
$SU(N_f)_L \otimes SU(N_f)_R/SU(N_f)_{L=R} = SU(N_f)$. 

In the standard model with $N_f = 2$ and arbitrary odd $N_c$ the electric 
charge of the up and down quarks is given by $Q = T^3_L + Y = 
T^3_L + T^3_R + \frac{1}{2} B$. Here $T^3_L$ and $T^3_R$ are the diagonal 
generators of $SU(2)_L \otimes SU(2)_R$ and $B$ is the baryon number. Since
$U(1)_B$, and hence $U(1)_Y$ and $U(1)_{em}$, are not subgroups of 
$SU(2)_L \otimes SU(2)_R$, it is not straightforward to gauge the electroweak
symmetry or even just electromagnetism in the low-energy pion effective theory.
It was first realized by Skyrme \cite{Sky61} that a baryon current can be 
constructed from pion fields, although the pions themselves do not carry any 
baryon number. In particular, there are solitons --- the so-called Skyrmions 
--- which indeed represent baryons. While the detailed structure of the 
Skyrmion (and even the question of its stability against shrinking) are beyond 
reach of chiral perturbation theory, the fact that pion field configurations 
with non-zero baryon number exist has profound consequences. In particular, if 
one wants to gauge the weak interactions at the level of the effective theory, 
one must quantize the Skyrmion as a fermion \cite{Wit83,DHo84} in order to 
cancel Witten's global anomaly. In order to account for the baryon number 
contribution $\frac{1}{2} B$ to the hypercharge $Y$ or the electric charge $Q$,
one must also include a gauge invariant modification of Skyrme's baryon number 
current --- the so-called Goldstone-Wilczek current \cite{Gol81} --- in the 
chiral Lagrangian. The Goldstone-Wilczek term cancels the triangle anomalies of
the lepton sector and is thus necessary to correctly describe the electroweak 
interactions of pions. In particular, the Goldstone-Wilczek current is 
responsible for the decay $\pi^0 \rightarrow \gamma \gamma$, which occurs 
because $Q_u + Q_d \neq 0$, i.e.\ because the quark charge matrix $Q$ is not a 
traceless generator of $SU(2)_{L=R}$. As we will see, for $N_f = 2$, the number
of colors $N_c$ does not appear explicitly in the low-energy effective theory 
of the standard model. Hence, it is then impossible to directly see the number 
of colors in low-energy experiments with just pions and photons.

In the $N_f \geq 3$ case the Wess-Zumino-Witten term \cite{Wes71,Wit83} arises
with a quantized prefactor. Hence, besides the more familiar low-energy 
parameters like $F_\pi$, the chiral Lagrangian also contains an integer-valued 
low-energy parameter. As first shown by Witten \cite{Wit83}, in QCD this 
parameter is the number of colors $N_c$. Since all low-energy parameters (like
$F_\pi$) have some implicit $N_c$-dependence, it is perhaps not too surprising 
that the integer-valued parameter also depends on $N_c$. However, in contrast 
to the $N_f = 2$ case, this means that $N_c$ explicitly enters the low-energy 
Goldstone boson theory for $N_f \geq 3$. For example, there is a strong 
interaction vertex that turns two kaons into three pions. This vertex is 
directly proportional to $N_c$. Hence, by scattering pions and kaons at low 
energies, one can indeed figure out the number of colors. Once one has 
appreciated the presence of the integer-valued low-energy parameter (with value
$N_c$), this should not be too surprising. In particular, one is not surprised 
that other vertices depend, for example, on the low-energy parameter $F_\pi$, 
which implicitly depends on $N_c$. After all, it is natural that strong 
interaction processes depend implicitly or explicitly on the number of colors. 
It is perhaps more surprising that electromagnetic probes can see $N_c$ in 
low-energy experiments.

When one gauges electromagnetism in an $N_f = 3$ theory with quark charges 
$Q_u = \frac{2}{3}$, $Q_d = Q_s = - \frac{1}{3}$ (and hence with 
$Q_u + Q_d + Q_s = 0$) the Wess-Zumino-Witten term gives a decay width of 
$\pi^0 \rightarrow \gamma \gamma$ proportional to $N_c^2$. However, according 
to eq.(\ref{charges}) this charge assignment implicitly assumes $N_c = 3$. For 
arbitrary odd $N_c$ one has $Q_u = (1/N_c + 1)/2$, $Q_d = Q_s = (1/N_c - 1)/2$,
and hence $Q_u + Q_d + Q_s = (3/N_c - 1)/2 \neq 0$. This means that for 
$N_c \neq 3$ the charge matrix is not a traceless generator of $SU(3)_{L=R}$.
Then, as for $N_f = 2$, one has to include a Goldstone-Wilczek term even in 
the $N_f = 3$ case. As we will see, this term cancels the contribution of the 
Wess-Zumino-Witten term to the decay $\pi^0 \rightarrow \gamma \gamma$, and 
leads to a width that is independent of the number of colors. This is indeed 
the correct result for the standard model with arbitrary odd $N_c$. Still, for 
$N_f \geq 3$ there are processes involving photons, pions, $\eta$-mesons and 
kaons that allow one to see the number of colors. In particular, the width of 
the decay $\eta \rightarrow \pi^+ \pi^- \gamma$ is proportional to $N_c^2$. 
This decay deserves to become the future textbook process that implies that 
there are indeed three colors in our world.

This paper is organized as follows. In section 2 we derive eq.(\ref{charges}) 
by canceling the anomalies in the standard model with $N_c$ colors. In section 
3 we discuss the low-energy pion effective theory for $N_f = 2$ and arbitrary 
odd $N_c$. Section 4 deals with the $N_f \geq 3$ case. In section 5 we show 
that in all photon-pion vertices the factor $N_c$ drops out, while it survives 
in some processes involving photons, pions, $\eta$-mesons and kaons. Finally, 
section 6 contains our conclusions.

\section{A Consistent Standard Model with Arbitrary Odd $N_c$}

In this section, following ref.\cite{Abb90}, we use gauge anomaly cancellation 
conditions to determine the electroweak charges of quarks in a standard model 
with an arbitrary number of colors $N_c$. We leave the Higgs, gauge, and 
lepton sectors unchanged, and only adjust the quark sector in order to achieve 
anomaly cancellation. We must pay attention to Witten's nonperturbative global 
anomaly, to the perturbative gauge triangle anomalies, as well as to the 
gravitational anomaly. As in the standard model at $N_c = 3$, we demand anomaly
cancellation separately for each generation of fermions, and discuss this in 
the context of the first generation.

In the lepton sector of the first generation we have an $SU(2)_L$ doublet
containing the left-handed neutrino and the left-handed electron, as well as 
two $SU(2)_L$ singlets: the right-handed neutrino and the right-handed 
electron. In the quark sector we have $N_c$ $SU(2)_L$ doublets containing the 
left-handed up and down quarks of different colors, as well as $2N_c$ $SU(2)_L$
singlets containing the right-handed up and down quarks. Hence, the fermion 
fields of one generation are
\begin{equation}
\left(\begin{array}{c} \nu_L(x) \\ e_L(x) \end{array}\right), \ \nu_R(x), \ 
e_R(x), \ \left(\begin{array}{c} u^i_L(x) \\ d^i_L(x) \end{array}\right), \ 
u^i_R(x), \ d^i_R(x), \ i \in \{1,2,...,N_c\}.
\end{equation}
The leptons are color singlets while the quarks are in the $N_c$-dimensional 
fundamental representation of $SU(N_c)$. We leave the $U(1)_Y$ hypercharge
assignments of the leptons unchanged, i.e.
\begin{equation}
Y_{\nu_L} = Y_{e_L} = - \frac{1}{2}, \quad Y_{\nu_R} = 0, \quad Y_{e_R} = - 1.
\end{equation}
Note that in the lepton sector
\begin{equation}
Y = T^3_R - \frac{1}{2} L,
\end{equation}
where $T^3_R$ is the diagonal generator of $SU(2)_R$ and $L$ is lepton number. 
The corresponding electric charges result from
\begin{equation}
Q_l = T^3_L + Y = T^3_L + T^3_R - \frac{1}{2} L,
\end{equation}
and are given by $Q_\nu = 0$ and $Q_e = - 1$. On the other hand, we leave the 
quark hypercharges $Y_{u_L} = Y_{d_L} = Y_L$, $Y_{u_R}$, and $Y_{d_R}$ as free 
parameters to be determined by the anomaly cancellation conditions.

In four space-time dimensions (compactified to the sphere $S^4$) the gauge 
transformations $L(x) \in SU(2)_L$ fall into two topologically distinct classes
characterized by a ``winding'' number $\mbox{Sign}[L] = \pm 1 \in 
\Pi_4[SU(2)_L] = \Z(2)$. Gauge transformations $L(x)$ that can be deformed 
continuously into the trivial gauge transformation have $\mbox{Sign}[L] = 1$, 
while all others have $\mbox{Sign}[L] = - 1$. The fermion determinant of a 
single $SU(2)_L$ doublet changes sign under a nontrivial $SU(2)_L$ gauge 
transformation with $\mbox{Sign}[L] = - 1$ and is thus not gauge invariant. In 
order to obtain a gauge invariant theory one hence needs an even number of 
$SU(2)_L$ doublets. Since there are one lepton doublet and $N_c$ quark 
doublets, $N_c$ must be odd in order to cancel Witten's global anomaly. This 
implies that the standard model is consistent only if the baryons are fermions.

In the next step we cancel the triangle anomalies which are proportional to
\begin{equation}
A^{abc} = \mbox{Tr}_L[(T^a T^b + T^b T^a)T^c] - 
\mbox{Tr}_R[(T^a T^b + T^b T^a)T^c].
\end{equation}
Here, the $T^a$ with $a \in \{1,2,3\}$ refer to the generators of $SU(2)_L$, 
$T^4 = Y$, and the $N_c^2 - 1$ remaining $T^a$, with 
$a - 4 \in \{1,2,...,N_c^2 - 1\}$, generate the color gauge group $SU(N_c)$. 
The traces are over the left- and right-handed fields, respectively. If two 
indices are color indices and the third index belongs to $U(1)_Y$ (i.e.\ 
$c = 4$), the anomaly cancels only if the quark hypercharges satisfy
\begin{equation}
\label{anomaly1}
2 Y_L - Y_{u_R} - Y_{d_R} = 0.
\end{equation}
When two indices belong to $SU(2)_L$ and the third index belongs to $U(1)_Y$ 
the anomaly cancellation condition involves both quarks and leptons and takes 
the form
\begin{equation}
\label{anomaly2}
2 N_c Y_L + Y_{\nu_L} + Y_{e_L} = 0 \ \Rightarrow \ Y_L = \frac{1}{2 N_c}.
\end{equation}
Finally, when all three indices belong to $U(1)_Y$, the anomaly $A^{444}$ 
vanishes if
\begin{equation}
\label{anomaly3}
N_c (2 Y_L^3 - Y_{u_R}^3 - Y_{d_R}^3) + 
Y_{\nu_L}^3 + Y_{e_L}^3 - Y_{\nu_R}^3 - Y_{e_R}^3 = 0 \ \Rightarrow \
2 Y_L^3 - Y_{u_R}^3 - Y_{d_R}^3 = - \frac{3}{4 N_c}.
\end{equation}
The cancellation of the gravitational anomaly also yields eq.(\ref{anomaly1})
and hence does not imply additional constraints. Combining the anomaly 
cancellation conditions eqs.(\ref{anomaly1},\ref{anomaly2},\ref{anomaly3}) one 
finally obtains
\begin{equation}
\label{hypercharges}
Y_L = \frac{1}{2 N_c}, \quad Y_{u_R} = \frac{1}{2}(\frac{1}{N_c} + 1), \quad
Y_{d_R} = \frac{1}{2}(\frac{1}{N_c} - 1).
\end{equation}
In the quark sector we can hence write
\begin{equation}
Y = T^3_R + \frac{1}{2} B,
\end{equation}
where $B = 1/N_c$ is the baryon number of a quark. The quark electric charges 
are given by
\begin{equation}
\label{quarkcharges}
Q_q = T^3_L + T^3_R + \frac{1}{2} B,
\end{equation}
which results in the up and down quark charges of eq.(\ref{charges}). The
general expression for the electric charge, valid for both quarks and leptons,
is
\begin{equation}
Q = T^3_L + T^3_R + \frac{1}{2} (B - L).
\end{equation}
Since $Q$ as well as $T^3_L + T^3_R$ generate symmetries of the standard model,
$B - L$ is a good symmetry as well. 

At this point we have constructed an anomaly free generalization of the 
standard model with an arbitrary odd number of colors. This shows explicitly 
that the consistency requirement of anomaly cancellation does not imply 
$N_c = 3$. Even the constraint that $N_c$ must be odd resulted only because we 
insisted that the anomalies cancel within a single generation. When there is an
even number of generations, the global anomaly cancels automatically, and $N_c$
could then as well be even. In that case, the baryons are bosons which, as a 
consequence of eq.(\ref{charges}), have half-integer charges. For odd $N_c$, on
the other hand, the baryons are fermions with integer electric charges.

It should be noted that after canceling the gauge anomalies, there are still 
anomalies in some global symmetries. For example, the baryon number current
\begin{equation}
j_\mu = \frac{1}{N_c} \sum_{i = 1}^{N_c}(\bar u^i_L \gamma_\mu u^i_L + 
\bar u^i_R \gamma_\mu u^i_R + \bar d^i_L \gamma_\mu d^i_L +
\bar d^i_R \gamma_\mu d^i_R)
\end{equation}
is not conserved due to the 't Hooft anomaly \cite{tHo76}. Its divergence is
given by
\begin{equation}
\label{bviolation}
\p_\mu j_\mu = - \frac{1}{32 \pi^2} \varepsilon_{\mu\nu\rho\sigma} \mbox{Tr}
[W_{\mu\nu} W_{\rho\sigma}] + \frac{1}{32 \pi^2} \varepsilon_{\mu\nu\rho\sigma}
\mbox{Tr}[B_{\mu\nu} B_{\rho\sigma}].
\end{equation}
Here $W_\mu = i g W_\mu^a T^a$ is the $SU(2)_L$ gauge field with gauge coupling
$g$ and field strength
\begin{equation}
W_{\mu\nu} = \p_\mu W_\nu - \p_\nu W_\mu + [W_\mu,W_\nu],
\end{equation}
and $B_\mu = i g' B_\mu^3 T^3$ is the $U(1)_Y$ gauge field with gauge coupling 
$g'$ and field strength
\begin{equation}
B_{\mu\nu} = \p_\mu B_\nu - \p_\nu B_\mu.
\end{equation}
The first term in eq.(\ref{bviolation}) results from a triangle diagram with an
internal quark loop attached to two external $SU(2)_L$ currents and one 
external baryon current. This diagram is proportional to
\begin{equation}
\label{triangle1}
\mbox{Tr}_L[(T^3)^2 B] = N_c \frac{1}{2} \frac{1}{N_c} = \frac{1}{2}.
\end{equation}
The second term in eq.(\ref{bviolation}) comes from a triangle diagram with an 
internal quark loop attached to two external $U(1)_Y$ currents and one external
baryon current. That diagram is proportional to
\begin{eqnarray}
\label{triangle2}
&&\mbox{Tr}_L[Y^2 B] - \mbox{Tr}_R[Y^2 B] = 
N_c [2 Y_L^2 - Y_{u_R}^2 - Y_{d_R}^2] \frac{1}{N_c} = \nonumber \\
&&\frac{1}{4}[2 \frac{1}{N_c^2} - (\frac{1}{N_c} + 1)^2 - 
(\frac{1}{N_c} - 1)^2] = - \frac{1}{2}.
\end{eqnarray}
It is interesting that both in eq.(\ref{triangle1}) and in eq.(\ref{triangle2})
the $N_c$-dependence cancels completely. Also, note that an electroweak 
instanton with topological charge
\begin{equation}
- \frac{1}{32 \pi^2} \int d^4x \ \varepsilon_{\mu\nu\rho\sigma} \mbox{Tr}
[W_{\mu\nu} W_{\rho\sigma}] = 1
\end{equation}
causes violation of baryon number conservation by one unit.

\section{Low-Energy Description of a Single Generation}

Let us first restrict ourselves to one generation of fermions. From the point
of view of the strong interactions, this is the $N_f = 2$ case of just up and
down quark flavors. Due to spontaneous chiral symmetry breaking from 
$SU(2)_L \otimes SU(2)_R$ to $SU(2)_{L=R}$, the low-energy degrees of freedom
are the pseudo-Goldstone pion fields
\begin{equation}
U(x) = \exp(2 i \pi^a(x) T^a/F_\pi),
\end{equation}
that live in the coset space 
$SU(2)_L \otimes SU(2)_R/SU(2)_{L=R} = SU(2)$. Note that we have introduced the
generators of $SU(2)$ such that $\mbox{Tr}(T^a T^b) = \frac{1}{2} \delta_{ab}$.
At low energies the pion dynamics is described by chiral perturbation theory. 
To lowest order, the Euclidean chiral perturbation theory action is given by
\cite{Gas85}
\begin{equation}
\label{action}
S[U] = \int d^4x \ \{\frac{F_\pi^2}{4} \mbox{Tr}[\p_\mu U^\dagger \p_\mu U] -
\frac{1}{4} \langle \bar\Psi \Psi \rangle \mbox{Tr}[{\cal M}(U + U^\dagger)]\}.
\end{equation}
Here $\langle \bar\Psi \Psi \rangle$ is the chiral condensate and 
${\cal M} = \mbox{diag}(m_u,m_d)$ is the quark mass matrix. For massless quarks
the action is invariant under global $SU(2)_L \otimes SU(2)_R$ transformations
\begin{equation}
U'(x) = L^\dagger U(x) R.
\end{equation}

The nontrivial homotopy group $\Pi_3[SU(2)] = \Z$ implies that, at every 
instant in time, the pion field is characterized by an integer winding number
\begin{equation}
B = \frac{1}{24 \pi^2} \int d^3x \ \varepsilon_{ijk} \mbox{Tr}
[(U^\dagger \p_i U)(U^\dagger \p_j U)(U^\dagger \p_k U)].
\end{equation}
Skyrme was first to suggest that $B$ should be identified with baryon number 
\cite{Sky61}. The baryon current
\begin{equation}
\label{bcurrent}
j_\mu = \frac{1}{24 \pi^2} \varepsilon_{\mu\nu\rho\sigma} \mbox{Tr}
[(U^\dagger \p_\nu U)(U^\dagger \p_\rho U)(U^\dagger \p_\sigma U)]
\end{equation}
is topologically conserved, i.e.\ $\p_\mu j_\mu = 0$.

The partition function of the pion field theory takes the form
\begin{equation}
\label{pathintegral}
Z = \int {\cal D}U \exp(- S[U]) \ \mbox{Sign}[U].
\end{equation}
The ``winding'' number $\mbox{Sign}[U]$ is an element of the nontrivial 
homotopy group $\Pi_4[SU(2)] = \Pi_4[S^3] = \Z(2)$. It can be identified as the
fermion permutation sign of the Skyrme soliton. For example, a pion field 
configuration $U$ in which two Skyrmions interchange their positions as they 
evolve in time has $\mbox{Sign}[U] = - 1$ \cite{Wit83}. When the Skyrmion is 
quantized as a fermion, $\mbox{Sign}[U]$ must be included in the pion path 
integral in order to correctly implement the Pauli principle for Skyrmions. 
Similarly, a configuration $U$ in which a single Skyrmion rotates by $2 \pi$ 
during its time evolution also has $\mbox{Sign}[U] = - 1$. The inclusion of 
$\mbox{Sign}[U]$ in the path integral therefore automatically ensures that the 
Skyrmion has half-integer spin. As we have seen in the previous section, the 
standard model is free of Witten's global anomaly only for odd $N_c$. Hence, 
the Skyrmions of the low-energy pion effective theory should be quantized as 
fermions. As we will see below, one indeed must include $\mbox{Sign}[U]$ in the
path integral in order to cancel the global anomaly also at the level of the 
effective theory \cite{Wit83,DHo84}.

In the next step we want to gauge both $SU(2)_L$ and $U(1)_Y$ in order to 
obtain a low-energy effective theory of the full standard model. In particular,
we are interested if $N_c$ enters the effective theory explicitly. Of course,
as discussed earlier, $N_c$ enters implicitly through $F_\pi$ and 
$\langle \bar\Psi \Psi \rangle$. 

Gauging $SU(2)_L$ is straightforward. First, one just replaces ordinary 
derivatives by covariant derivatives
\begin{equation}
D_\mu U = (\p_\mu + W_\mu) U.
\end{equation}
Second, one replaces the quark mass matrix ${\cal M}$ by a coupling to the
standard model Higgs field which can be expressed as a matrix
\begin{equation}
\Phi(x) = \pmatrix{\Phi_0^*(x) & \Phi_+(x) \cr - \Phi_+^*(x) & \Phi_0(x)}.
\end{equation}
The action then takes the form
\begin{equation}
\label{gaugedaction}
S[U,\Phi,W_\mu] = \int d^4x \ \{\frac{F_\pi^2}{4} 
\mbox{Tr}[D_\mu U^\dagger D_\mu U] - \frac{1}{4} \langle \bar\Psi \Psi \rangle 
\mbox{Tr}[{\cal F}^\dagger \Phi^\dagger U + U^\dagger \Phi {\cal F}]\},
\end{equation}
which is invariant under local transformations
\begin{equation}
U'(x) = L^\dagger(x) U(x), \quad \Phi'(x) = L^\dagger(x) \Phi(x), \quad
W_\mu'(x) = L^\dagger(x)(W_\mu(x) + \p_\mu) L(x).
\end{equation}
In the vacuum the Higgs field has the expectation value $\Phi(x) = v \1$, and
the up and down quark masses are obtained from the matrix of Yukawa couplings 
${\cal F} = \mbox{diag}(f_u,f_d)$ as
\begin{equation}
{\cal M} = {\cal F} v \ \Rightarrow \ m_u = f_u v, \quad m_d = f_d v.
\end{equation}

When one performs an $SU(2)_L$ gauge transformation $L(x)$ with 
$\mbox{Sign}[L] = - 1$ in the standard model, the fermion determinant of the 
leptons changes sign and is thus not gauge invariant. The global anomaly is 
canceled by an odd number of minus signs due to the $N_c$ quark $SU(2)_L$
doublets. At the level of the low-energy effective theory, the leptons are
still present but the quarks have been replaced by pion fields. Hence, the
question arises how the cancellation of the global anomaly is achieved at the
level of the low-energy effective theory. While the pion action 
$S[U,\Phi,W_\mu]$ of eq.(\ref{gaugedaction}) is gauge invariant, the path 
integral as a whole is not. This is because
\begin{equation}
\mbox{Sign}[U'] = \mbox{Sign}[L U] = \mbox{Sign}[L] \ \mbox{Sign}[U].
\end{equation}
Hence, as pointed out by Witten \cite{Wit83} and by D'Hoker and Farhi 
\cite{DHo84}, the $SU(2)_L$ gauge variation of the fermion permutation sign of 
the Skyrmions cancels the global anomaly of the leptons. In this way, the 
effective theory inherits the global anomaly cancellation constraint that $N_c$
must be odd. The low-energy theory is gauge invariant only if its baryons are 
quantized as fermions.

At the quark level, we have seen that the hypercharge $Y = T^3_R + \frac{1}{2} 
B$ contains the baryon number $B$. Since $U(1)_B$ is not a subgroup of 
$SU(2)_L \otimes SU(2)_R$, it is not at all straightforward to gauge $U(1)_Y$. 
Gauging the $SU(2)_R$ component of $Y$ simply amounts to extending the 
covariant derivative to
\begin{equation}
\label{covariant}
D_\mu U = \p_\mu U + W_\mu U - U B_\mu,
\end{equation}
where $B_\mu$ is the $U(1)_Y$ gauge field.
However, incorporating the covariant derivatives alone is not sufficient to 
correctly gauge $U(1)_Y$. Although the pions themselves do not carry baryon 
number, it is still important to incorporate the baryon current in the 
effective theory since $Y$ contains the baryon number $B$. In particular, if 
one would not include the baryon current, the decay $\pi^0 \rightarrow \gamma 
\gamma$ would not happen in the effective theory.

Of course, when $SU(2)_L$ is gauged, baryon number conservation is violated by 
electroweak instantons according to eq.(\ref{bviolation}). When one replaces 
ordinary derivatives by covariant ones in the baryon current of 
eq.(\ref{bcurrent}), its divergence does not obey eq.(\ref{bviolation}).
Instead, one should consider the Goldstone-Wilczek current \cite{Gol81,DHo84}
\begin{eqnarray}
\label{GWcurrent}
j^{GW}_\mu&=&\frac{1}{24 \pi^2} \varepsilon_{\mu\nu\rho\sigma} \mbox{Tr}
[(U^\dagger D_\nu U)(U^\dagger D_\rho U)(U^\dagger D_\sigma U)] \nonumber \\
&-&\frac{1}{16 \pi^2} \varepsilon_{\mu\nu\rho\sigma} \mbox{Tr}
[W_{\nu\rho} (D_\sigma U U^\dagger)] 
- \frac{1}{16 \pi^2} \varepsilon_{\mu\nu\rho\sigma} \mbox{Tr}
[B_{\nu\rho} (U^\dagger D_\sigma U)],
\end{eqnarray}
whose divergence is indeed given by
\begin{equation}
\p_\mu j^{GW}_\mu = - \frac{1}{32 \pi^2} \varepsilon_{\mu\nu\rho\sigma} 
\mbox{Tr}[W_{\mu\nu} W_{\rho\sigma}] +
\frac{1}{32 \pi^2} \varepsilon_{\mu\nu\rho\sigma} 
\mbox{Tr}[B_{\mu\nu} B_{\rho\sigma}].
\end{equation}

Since, at the quark level, the $U(1)_Y$ gauge field couples to baryon number
with strength $g'/2$, the same must be true in the effective theory. This
implies that, in order to gauge $U(1)_Y$ correctly, one must include the
Goldstone-Wilczek current in the low-energy effective action. This is achieved
by introducing a Goldstone-Wilczek term
\begin{equation}
S_{GW}[U,W_\mu,B_\mu] = \frac{g'}{2} \int d^4x \ B_\mu^3 j_\mu^{GW}.
\end{equation}
It should be noted that this term alone is not gauge invariant. In this respect
it is similar to the term $\mbox{Sign}[U]$. While $\mbox{Sign}[U]$ varies under
topologically nontrivial $SU(2)_L$ gauge transformations, the gauge variance is
exactly what one needs to cancel the global anomaly in the lepton sector. 
Similarly, while $j^{GW}_\mu$ is both $SU(2)_L$ and $U(1)_Y$ gauge invariant, 
$S_{GW}[U,W_\mu,B_\mu]$ is only $SU(2)_L$ gauge invariant, but varies under 
$U(1)_Y$ gauge transformations $B'^3_\mu = B^3_\mu + \p_\mu \varphi$. The 
violation of gauge invariance is determined by
\begin{eqnarray}
&&S_{GW}[U',W_\mu,B_\mu'] - S_{GW}[U,W_\mu,B_\mu] = \int d^4x \ \p_\mu \varphi 
j_\mu^{GW} = - \int d^4x \ \varphi \p_\mu j_\mu^{GW} \nonumber \\
&&=\int d^4x \ \varphi \{\frac{1}{32 \pi^2} \varepsilon_{\mu\nu\rho\sigma} 
\mbox{Tr}[W_{\mu\nu} W_{\rho\sigma}] - \frac{1}{32 \pi^2} 
\varepsilon_{\mu\nu\rho\sigma} \mbox{Tr}[B_{\mu\nu} B_{\rho\sigma}]\}.
\end{eqnarray}
This gauge variation is exactly what is needed to cancel the triangle anomalies
in the lepton sector and render the whole theory gauge invariant. 

The path integral of pions in the background of Higgs and gauge fields finally
takes the form
\begin{equation}
Z[\Phi,W_\mu,B_\mu] = \int {\cal D}U \exp(- S[U,\Phi,W_\mu,B_\mu]) \
\mbox{Sign}[U] \exp(i S_{GW}[U,W_\mu,B_\mu]).
\end{equation}
Nowhere in this expression does $N_c$ appear as an explicit parameter. It only
appears implicitly in $S[U,\Phi,W_\mu,B_\mu]$ through parameters like $F_\pi$ 
and $\langle \bar\Psi \Psi \rangle$. It should be noted that 
$S[U,\Phi,W_\mu,B_\mu]$ contains all normal parity contributions to the 
effective action, not only the leading terms given in eq.(\ref{action}). The 
Goldstone-Wilczek term, on the other hand, contains the anomalous parity
contributions.

Let us also discuss how to gauge just $U(1)_{em}$. According to 
eq.(\ref{quarkcharges}), at the quark level the electric charge is given by 
$Q = T^3_L + T^3_R + \frac{1}{2} B$. This implies that the electromagnetic 
covariant derivative takes the form
\begin{equation}
D_\mu U = \p_\mu U + i e A_\mu [T^3,U].
\end{equation}
This is consistent with eq.(\ref{covariant}) because the photon and $Z^0$ boson
fields are related to $W_\mu^3$ and $B_\mu^3$ by
\begin{equation}
W_\mu^3 = \frac{g' A_\mu + g Z_\mu}{\sqrt{g^2 + g'^2}}, \
B_\mu^3 = \frac{g A_\mu - g' Z_\mu}{\sqrt{g^2 + g'^2}},
\end{equation}
and the electric charge is given by
\begin{equation}
e = \frac{g g'}{\sqrt{g^2 + g'^2}}.
\end{equation}
The Goldstone-Wilczek current now takes the form
\begin{eqnarray}
\label{GWcurrentem}
j^{GW}_\mu&=&\frac{1}{24 \pi^2} \varepsilon_{\mu\nu\rho\sigma} \mbox{Tr}
[(U^\dagger D_\nu U)(U^\dagger D_\rho U)(U^\dagger D_\sigma U)] \nonumber \\
&-&\frac{i e}{16 \pi^2} \varepsilon_{\mu\nu\rho\sigma} F_{\nu\rho} \mbox{Tr}
[T^3 (D_\sigma U U^\dagger + U^\dagger D_\sigma U)].
\end{eqnarray}
In this case, the theory is vector-like and the baryon current is conserved.
The path integral then takes the form
\begin{equation}
Z[A_\mu] = \int {\cal D}U \exp(- S[U,A_\mu]) \ \mbox{Sign}[U]
\exp(i S_{GW}[U,A_\mu]),
\end{equation}
with
\begin{equation}
\label{SGWem}
S_{GW}[U,A_\mu] = \frac{e}{2} \int d^4x \ A_\mu j_\mu^{GW}.
\end{equation}

One can now identify the vertex responsible for the decay 
$\pi^0 \rightarrow \gamma \gamma$. Putting $U \approx 1 + 2 i \pi^0 T^3/F_\pi$,
after partial integration the second term in the Goldstone-Wilczek current of 
eq.(\ref{GWcurrentem}) indeed yields the vertex
\begin{equation}
{\cal L}_{\pi^0 \gamma \gamma} = -i \frac{e^2}{32 \pi^2 F_\pi} \pi^0 
\varepsilon_{\mu\nu\rho\sigma} F_{\mu\nu} F_{\rho\sigma}.
\end{equation}
By now it should no longer come as a surprise that this vertex is not 
proportional to $N_c$. 

Another vertex that is often claimed to be proportional to $N_c$ is 
${\cal L}_{\pi^0 \pi^+ \pi^- \gamma}$. In the microscopic theory this vertex 
results from a quark box diagram with three external pion and one external 
photon line. This diagram is proportional to
\begin{equation}
\mbox{Tr}(T^3 [T^+,T^-] Q) = \frac{N_c}{4} (Q_u + Q_d),
\end{equation}
where $T^\pm = (T^1 \pm i T^2)/\sqrt{2}$. If one uses $Q_u = \frac{2}{3}$ and 
$Q_d = - \frac{1}{3}$ independent of $N_c$, this expression is obviously 
proportional to $N_c$. However, if one uses the consistent quark charges of 
eq.(\ref{charges}) it becomes $N_c$-independent because
\begin{equation}
\frac{N_c}{4}(Q_u + Q_d) = \frac{N_c}{4}[\frac{1}{2}(\frac{1}{N_c} + 1) + 
\frac{1}{2}(\frac{1}{N_c} - 1)] = \frac{1}{4}.
\end{equation}
Using $U \approx 1 + 2 i \pi^a T^a/F_\pi$ as well as $\pi^\pm = 
(\pi^1 \pm i \pi^2)/\sqrt{2}$, in the anomaly free standard model with
consistent quark charges this vertex is
\begin{equation}
{\cal L}_{\pi^0 \pi^+ \pi^- \gamma} = \frac{e}{4 \pi^2 F_\pi^3}
\varepsilon_{\mu\nu\rho\sigma} A_\mu \p_\nu \pi^0 \p_\rho \pi^+ \p_\sigma 
\pi^-,
\end{equation}
which is again $N_c$-independent. The same is true for any other photon-pion
vertex. Hence, contrary to common lore, it is impossible to literally see $N_c$
in experiments with just photons and pions.

\section{Low-Energy Description of $N_f \geq 3$ Flavors}

In a world with $N_f \geq 3$ massless quarks the chiral symmetry is 
$SU(N_f)_L \otimes SU(N_f)_R$ which is spontaneously broken to $SU(N_f)_{L=R}$.
Consequently, the Goldstone boson fields now live in the coset space
$SU(N_f)_L \otimes SU(N_f)_R/SU(N_f)_{L=R} = SU(N_f)$. The leading order chiral
perturbation theory action takes the same form of eq.(\ref{action}) as in the
$N_f = 2$ case. Since $\Pi_3[SU(N_f)] = \Z$ for any $N_f \geq 3$, the Skyrme 
and Goldstone-Wilczek currents of eqs.(\ref{bcurrent},\ref{GWcurrent}) also 
still have the same form.

However, in contrast to the $N_f = 2$ case, $\Pi_4[SU(N_f)] = \{0\}$ for 
$N_f \geq 3$. Consequently, any space-time dependent Goldstone boson field 
$U(x) \in SU(N_f)$ can now be continuously deformed into the trivial field 
$U(x) = \1$. By introducing a fifth coordinate $x_5 \in [0,1]$ which plays the 
role of a deformation parameter, one can extend the 4-dimensional field $U(x)$ 
to a field $U(x,x_5)$ on the 5-dimensional hemisphere $H^5$ whose boundary 
$\p H^5 = S^4$ is (compactified) space-time, such that $U(x,0) = \1$ and 
$U(x,1) = U(x)$. This allows one to write down the Wess-Zumino-Witten term 
\cite{Wit83} with the action
\begin{equation}
\label{WZW}
S_{WZW}[U] = \frac{1}{240 \pi^2 i} \int_{H^5} d^5x \ 
\varepsilon_{\mu\nu\rho\sigma\lambda} \mbox{Tr}[(U^\dagger \p_\mu U)
(U^\dagger \p_\nu U)(U^\dagger \p_\rho U)(U^\dagger \p_\sigma U)
(U^\dagger \p_\lambda U)].
\end{equation}
Note that the factor $i$ in eq.(\ref{WZW}) is necessary in order to get a 
real-valued result. Of course, the 4-dimensional physics should be independent 
of how the field $U(x,x_5)$ is deformed into the bulk of the fifth dimension. 
It should only depend on the boundary values $U(x)$, i.e.\ on the Goldstone 
boson field in the physical part of space-time. This is possible because the 
integrand in eq.(\ref{WZW}) is a total divergence. In particular, $S_{WZW}[U]$ 
is closely related to the winding number $\Pi_5[SU(N_f)] = \Z$. In fact, if the
integration in eq.(\ref{WZW}) were performed over a sphere $S^5$ instead of the
hemisphere $H^5$, the result would be $2 \pi$ times the integer winding number 
of $U(x,x_5)$. Hence, modulo integers, $S_{WZW}[U]$ gets contributions only 
from the boundary of $H^5$, i.e.\ from the 4-dimensional physical space-time 
$S^4$. Of course, one must still ensure that the integer contribution from the 
5-dimensional bulk cancels. This is indeed the case, because $S_{WZW}[U]$ 
enters the path integral,
\begin{equation}
\label{PI>2}
Z = \int {\cal D}U \exp(- S[U]) \exp(i N_c S_{WZW}[U]),
\end{equation}
with a quantized prefactor --- the number of colors $N_c$. It should be noted 
that eq.(\ref{PI>2}) is the natural extension of eq.(\ref{pathintegral}) in the
$N_f = 2$ case. In fact, one can show that for $U(x) \in SU(2)$
\begin{equation}
\exp(i N_c S_{WZW}[U]) = \mbox{Sign}[U]^{N_c}.
\end{equation}
The argument of the Wess-Zumino-Witten term is a 5-dimensional Goldstone boson
field $U(x,x_5) \in SU(N_f)$ which reduces to a 4-dimensional $SU(2)$ field
$U(x)$ at the boundary of $H^5$. The argument of the sign factor, on the other
hand, is just the 4-dimensional field $U(x) \in SU(2)$. Indeed, in the 
$N_f \geq 3$ theory, the Wess-Zumino-Witten term $\exp(i N_c S_{WZW}[U])$
plays a similar role as $\mbox{Sign}[U]$ in the $N_f = 2$ case. In particular, 
for odd $N_c$ it ensures that the Skyrmion is again quantized as a fermion with
half-integer spin \cite{Wit83}. It also ensures that the global anomaly is 
properly canceled when one gauges $SU(2)_L$. In addition, for $N_f \geq 3$ the 
Wess-Zumino-Witten term also contributes to effects that are entirely due to 
the Goldstone-Wilczek term in the $N_f = 2$ case. For example, it also
contributes to the triangle anomalies and to the $\pi^0$ decay.

Unlike in the $N_f = 2$ case, for $N_f \geq 3$ the Wess-Zumino-Witten term also
breaks the unwanted intrinsic parity symmetry $P_0$ that is present in the 
Goldstone boson action $S[U]$, but not in QCD \cite{Wit83}. The full parity 
operation $P$, of course, is a symmetry of QCD, at least for vanishing vacuum 
angle $\theta = 0$. Parity acts on the pseudo-scalar Goldstone bosons 
$\pi^a(\vec x,t)$ by spatial inversion accompanied by a sign-change, i.e.\ 
$^P\pi^a(\vec x,t) = - \pi^a(- \vec x,t)$. For the field $U$ the full parity 
transformation $P$ takes the form
\begin{equation}
^PU(\vec x,t) = U^\dagger(- \vec x,t),
\end{equation}
while the intrinsic parity $P_0$ leaves out the spatial inversion and thus
takes the form
\begin{equation}
^{P_0}U(\vec x,t) = U^\dagger(\vec x,t).
\end{equation}
If $P_0$ were a symmetry of QCD, the number of Goldstone bosons would be
conserved modulo two, i.e.\ no strong interaction process could change the 
number of Goldstone bosons from even to odd. For $N_f = 2$ this is indeed the 
case, and $P_0$ is, in fact, nothing but $G$-parity \cite{Lee56}. For 
$N_f \geq 3$, on the other hand, intrinsic parity is not a symmetry of QCD. For
example, the $\phi$-meson decays both into two kaons and into three pions. 
However, the Goldstone boson action $S[U]$ is indeed invariant under $P_0$,
\begin{equation}
S[^{P_0}U] = S[U^\dagger] = S[U],
\end{equation}
and hence has more symmetry than the underlying QCD action. The 
Wess-Zumino-Witten action, on the other hand, is odd under $P_0$,
\begin{equation}
S_{WZW}[^{P_0}U] = S_{WZW}[U^\dagger] = - S_{WZW}[U],
\end{equation}
and thus reduces the symmetry of the effective theory to the one of QCD.

It is remarkable that an integer parameter (with value $N_c$) appears 
explicitly in the low-energy effective theory of QCD with $N_f \geq 3$. In 
particular, this means that some low-energy processes involving more than two 
flavors indeed depend directly on how many quarks there are inside a proton. 
For example, there is a vertex in the Wess-Zumino-Witten term that turns two 
kaons into three pions. This vertex is directly proportional to $N_c$. When one
gauges $SU(2)_L$ and $U(1)_Y$, or even just $U(1)_{em}$, the explicit $N_c$ 
factor also affects some electroweak processes. For example, by gauging 
$U(1)_{em}$ in an $N_f = 3$ theory with the quark charge matrix
\begin{equation}
\label{Qreal}
Q' = \mbox{diag}(\frac{2}{3},- \frac{1}{3},- \frac{1}{3}),
\end{equation} 
Witten has shown that the vertex ${\cal L}_{\pi^0 \gamma \gamma}$ is 
proportional to $N_c$ \cite{Wit83}. While this is the correct answer if one 
varies $N_c$ without adjusting the quark charges, it is not what one gets for 
the anomaly free standard model with arbitrary odd $N_c$. In that case, the 
appropriate quark charge matrix is given by
\begin{eqnarray}
\label{Qmatrix}
Q&=&\mbox{diag}(Q_u,Q_d,Q_s) = 
\mbox{diag}(\frac{1}{2}(\frac{1}{N_c} + 1),\frac{1}{2}(\frac{1}{N_c} - 1),
\frac{1}{2}(\frac{1}{N_c} - 1)) \nonumber \\
&=&Q' + (1 - \frac{N_c}{3}) \frac{1}{2} B,
\end{eqnarray}
and hence for $N_c \neq 3$ it is not a generator of $SU(3)_{L=R}$. This implies
that, like in the $N_f = 2$ case, we must also include a Goldstone-Wilczek term
in the $N_f = 3$ effective theory. However, in the $N_f = 3$ case, the 
Goldstone-Wilczek term is a factor of $(1 - N_c/3)$ larger than in the 
$N_f = 2$ case and we thus obtain
\begin{equation}
\label{ZNf3}
Z[A_\mu] = \int {\cal D}U \exp(- S[U,A_\mu]) \exp(i N_c S_{WZW}[U,A_\mu])
\exp(i (1 - \frac{N_c}{3}) S_{GW}[U,A_\mu]).
\end{equation}

The $U(1)_{em}$-gauged Wess-Zumino-Witten term takes the form 
\cite{Wit83,Cho84,Kaw84,Man85}
\begin{eqnarray}
S_{WZW}[U,A_\mu]&=&S_{WZW}[U] + \frac{e}{48 \pi^2} \int d^4x \ 
\varepsilon_{\mu\nu\rho\sigma} A_\mu \nonumber \\ &\times&
\mbox{Tr}\{Q'[(\p_\nu U U^\dagger)(\p_\rho U U^\dagger)(\p_\sigma U U^\dagger) 
+ (U^\dagger \p_\nu U)(U^\dagger \p_\rho U)(U^\dagger \p_\sigma U)]\}
\nonumber \\
&-&\frac{i e^2}{48 \pi^2} \int d^4x \ \varepsilon_{\mu\nu\rho\sigma} A_\mu
F_{\nu\rho} \mbox{Tr}\{Q'(\p_\sigma U U^\dagger)
[Q' + \frac{1}{2} U Q' U^\dagger] \nonumber \\
&+&Q'(U^\dagger \p_\sigma U)[Q' + \frac{1}{2} U^\dagger Q' U]\}.
\end{eqnarray}
In the literature the two terms containing $U Q' U^\dagger$ and 
$U^\dagger Q' U$ are sometimes replaced by a single term \cite{Wit83,Cho84}. 
Our analysis agrees with the one in \cite{Kaw84}. We also write down the 
Goldstone-Wilczek term for general $N_f$
\begin{eqnarray}
S_{GW}[U,A_\mu]&=&\frac{e}{48 \pi^2} \int d^4x \ 
\varepsilon_{\mu\nu\rho\sigma} A_\mu 
\mbox{Tr}[(U^\dagger \p_\nu U)(U^\dagger \p_\rho U)(U^\dagger \p_\sigma U)] 
\nonumber \\
&-&\frac{i e^2}{32 \pi^2} \int d^4x \ \varepsilon_{\mu\nu\rho\sigma} A_\mu
F_{\nu\rho} \mbox{Tr}[Q'(\p_\sigma U U^\dagger + U^\dagger \p_\sigma U)].
\end{eqnarray}
Note that all derivatives in this expression can be replaced by covariant
derivatives without changing the result.

According to eq.(\ref{ZNf3}), for $N_c = 3$ no Goldstone-Wilczek term arises, 
and the contribution from the Wess-Zumino-Witten term alone gives the full 
answer. In that case, the quark charge matrix of eq.(\ref{Qreal}) that was used
by Witten, is indeed the one of the consistent standard model. The 
Wess-Zumino-Witten term alone contributes a vertex 
${\cal L}_{\pi^0 \gamma \gamma}$ that is proportional to
\begin{equation}
N_c (Q_u'^2 - Q_d'^2) = \frac{N_c}{3},
\end{equation}
and hence a factor $N_c/3$ stronger than the correct vertex in the consistent 
standard model for arbitrary odd $N_c$. However, for $N_c \neq 3$, there is 
also the Goldstone-Wilczek term, which contributes $(1 - N_c/3)$ times the 
correct vertex that was already obtained in the $N_f = 2$ case. The 
$N_c$-dependent part of the Goldstone-Wilczek term completely cancels the
contribution of the Wess-Zumino-Witten term, and hence the $N_c$-independent 
part alone indeed gives the correct strength
\begin{equation}
\frac{N_c}{3} + (1 - \frac{N_c}{3}) = 1.
\end{equation}
The above consideration can be trivially extended to a general number $N_f$ of 
light flavors. Let us consider the consistent standard model with arbitrary odd
$N_c$ and with several generations of fermions. We assume that there are 
$N_u \geq 1$ light up-type quarks (up, charm, top) with charge 
$Q_u = (1/N_c + 1)/2$ and $N_d \geq 1$ light down-type quarks (down, strange, 
bottom) with charge $Q_d = (1/N_c - 1)/2$. In this case, $N_f = N_u + N_d$ and 
the charge matrix of the light quarks takes the form
\begin{equation}
Q = Q' + (1 - N_c \frac{N_d - N_u}{N_f}) \frac{1}{2} B,
\end{equation}
where $Q'$ is a traceless diagonal generator of $SU(N_f)_{L=R}$. In the 
corresponding low-energy effective theory,
\begin{eqnarray}
Z[A_\mu]&=&\int {\cal D}U \exp(- S[U,A_\mu]) \exp(i N_c S_{WZW}[U,A_\mu])
\nonumber \\
&\times&\exp(i (1 - N_c \frac{N_d - N_u}{N_f}) S_{GW}[U,A_\mu]),
\end{eqnarray}
there is a Goldstone-Wilczek current which contributes 
$(1 - N_c (N_d - N_u)/N_f)$ times the correct vertex 
${\cal L}_{\pi^0 \gamma \gamma}$. The Wess-Zumino-Witten term, on the other 
hand, is now coupled to the photon field by using the remaining $SU(N_f)_{L=R}$
generator $Q'$, which has diagonal elements $Q'_u$ and $Q'_d$ with
\begin{equation}
Q'_u = Q_u - \frac{1}{2}(\frac{1}{N_c} - \frac{N_d - N_u}{N_f}) =
\frac{N_d}{N_f}, \quad
Q'_d = Q_d - \frac{1}{2}(\frac{1}{N_c} - \frac{N_d - N_u}{N_f}) =
- \frac{N_u}{N_f}.
\end{equation}
The strength of the vertex ${\cal L}_{\pi^0 \gamma \gamma}$ that results from 
the Wess-Zumino-Witten term is given by
\begin{equation}
N_c (Q_u'^2 - Q_d'^2) = N_c \frac{N_d^2 - N_u^2}{N_f^2} = 
N_c \frac{N_d - N_u}{N_f}.
\end{equation}
As before, the $N_c$-dependent part of the Goldstone-Wilczek term completely 
cancels the contribution from the Wess-Zumino-Witten term, and the 
$N_c$-independent part of the Goldstone-Wilczek term gives the correct strength
\begin{equation}
N_c \frac{N_d - N_u}{N_f} + (1 - N_c \frac{N_d - N_u}{N_f}) = 1.
\end{equation}

\section{How Can One See $N_c$ for $N_f \geq 3$ ?}

Since the decay $\pi^0 \rightarrow \gamma \gamma$ does not allow one to see 
$N_c$, we now ask if other processes do. First, we consider only photons and 
pions (but no kaons or $\eta$-mesons) by embedding a pion $SU(2)$ sub-matrix 
$\tilde U(x)$ into $SU(N_f)$
\begin{equation}
U(x) = \pmatrix{\tilde U(x) & 0 \cr 0 & \1},
\end{equation}
where $\1$ is the $(N_f - 2) \times (N_f - 2)$ unit matrix. Then only a
$2 \times 2$ sub-matrix $\tilde Q'$ of the full $N_f \times N_f$ matrix $Q'$
enters the calculation. Using
\begin{equation}
\tilde Q' = \pmatrix{Q'_u & 0 \cr 0 & Q'_d} = \frac{N_d - N_u}{2 N_f} + T^3, \
\tilde {Q'}^2 = \pmatrix{{Q'_u}^2 & 0 \cr 0 & {Q'_d}^2} = 
\frac{N_d^2 + N_u^2}{2 N_f^2} + \frac{N_d - N_u}{N_f} T^3,
\end{equation}
it is straightforward to show that
\begin{equation}
N_c (S_{WZW}[U,A_\mu] - S_{WZW}[U]) + 
(1 - N_c \frac{N_d - N_u}{N_f}) S_{GW}[U,A_\mu] = S_{GW}[\tilde U,A_\mu],
\end{equation}
where $S_{GW}[\tilde U,A_\mu]$ is the $N_c$-independent $N_f = 2$ result of
eq.(\ref{SGWem}). This shows that the $N_f \geq 3$ result is fully consistent 
with the $N_f = 2$ calculation. In particular, all photon-pion vertices 
contained in the Wess-Zumino-term are completely canceled by the 
$N_c$-dependent piece of the Goldstone-Wilczek term. Hence, as we concluded
before, one cannot see $N_c$ directly in low-energy experiments of photons and
pions alone.

Next, we consider processes involving just photons and $\eta$-mesons in a 
theory with $N_u = 1$ and $N_d = 2$ (and hence $N_f = N_u + N_d = 3$). In that 
case, we write
\begin{equation}
U(x) = \exp(2 i \eta^8 T^8/F_\pi),
\end{equation}
with $T^8 = \frac{1}{2} \lambda^8$, and we use $Q' = T^3 + T^8/\sqrt{3}$. Then 
it is again straightforward to show that
\begin{eqnarray}
&&N_c (S_{WZW}[U,A_\mu] - S_{WZW}[U]) + 
(1 - N_c \frac{N_d - N_u}{N_f}) S_{GW}[U,A_\mu]
\nonumber \\
&&= N_c (S_{WZW}[U,A_\mu] - S_{WZW}[U]) + 
(1 - \frac{N_c}{3}) S_{GW}[U,A_\mu] 
\nonumber \\
&&= \frac{e^2}{32 \sqrt{3} \pi^2 F_\pi} \int d^4x \ \eta^8
\varepsilon_{\mu\nu\rho\sigma} F_{\mu\nu} F_{\rho\sigma},
\end{eqnarray}
which again does not contain $N_c$ explicitly. In the microscopic theory this 
vertex results from a quark triangle diagram that is proportional to
\begin{equation}
\mbox{Tr}(T^8 Q^2) = \frac{N_c}{2 \sqrt{3}}(Q_u^2 - Q_d^2) = 
\frac{1}{2 \sqrt{3}}.
\end{equation}

It should be noted that the physical $\eta$-meson is a mixture of the flavor 
octet $\eta^8$ and a flavor singlet $\eta^1$. In order to properly discuss the 
mixing, one must hence also include the $\eta^1$ field in the chiral 
Lagrangian. The combination of $\eta^1$ and $\eta^8$ orthogonal to the physical
$\eta$-meson, is the $\eta'$-meson. Indeed, in the large $N_c$ limit the 
$\eta'$-meson also becomes a Goldstone boson and should be included in the 
chiral Lagrangian \cite{DiV80,Wit80,Kai00}. Anomalous decays of Goldstone 
bosons and, in particular, the issue of meson mixing are discussed in 
\cite{Bij93}. The quark triangle diagram describing the decay of the flavor 
singlet $\eta^1 \rightarrow \gamma \gamma$ in the microscopic theory is 
proportional to
\begin{equation}
\mbox{Tr}(\frac{\1}{\sqrt{6}} Q^2) = \frac{N_c}{\sqrt{6}} (Q_u^2 + 2 Q_d^2) =
\frac{3}{4 \sqrt{6}}(N_c + \frac{1}{N_c} - \frac{2}{3}), 
\end{equation}
which is $N_c$-dependent. Hence, due to mixing, the width of the decay
$\eta \rightarrow \gamma \gamma$ indeed depends explicitly on $N_c$. However, 
this dependence is not so simple, because it is influenced by the amount of 
mixing which itself implicitly depends on $N_c$. Therefore, this decay is not 
too well suited for replacing the misleading textbook example 
$\pi^0 \rightarrow \gamma \gamma$ for providing experimental support for three
colors.

In the next step we consider the interactions of photons, pions, $\eta$-mesons
and kaons, again using $N_u = 1$, $N_d = 2$ and $N_f = 3$. In this case, it is 
easier to literally see $N_c$. In particular, the vertex
\begin{equation}
{\cal L}_{\eta^8 \pi^+ \pi^- \gamma} = \frac{e N_c}{4 \sqrt{3} \pi^2 F_\pi^3}
\varepsilon_{\mu\nu\rho\sigma} A_\mu \p_\nu \eta^8 \p_\rho \pi^+ \p_\sigma 
\pi^-,
\end{equation}
is proportional to $N_c$. It is interesting to note that only the 
Wess-Zumino-Witten term contributes to this process. Hence, in this case, there
is no cancellation with the $N_c$-dependent part of the Goldstone-Wilczek term.
In the microscopic theory this process results from a quark box diagram that is
proportional to
\begin{equation}
\mbox{Tr}(T^8 [T^+,T^-] Q) = \frac{N_c}{4 \sqrt{3}} (Q_u - Q_d) = 
\frac{N_c}{4 \sqrt{3}}.
\end{equation}
Because of mixing, one must also consider the corresponding vertex for the 
flavor singlet $\eta^1$. The interactions of photons with three Goldstone 
bosons and the effect of meson mixing on the decay 
$\eta \rightarrow \pi^+ \pi^- \gamma$ have been discussed in 
\cite{Bij93,Bij90}. The quark box diagram describing the decay of the flavor 
singlet $\eta^1 \rightarrow \pi^+ \pi^- \gamma$ is proportional to
\begin{equation}
\mbox{Tr}(\frac{\1}{\sqrt{6}} [T^+,T^-] Q) = \frac{N_c}{2 \sqrt{6}} 
(Q_u - Q_d) = \frac{N_c}{2 \sqrt{6}}. 
\end{equation}
Since both vertices ${\cal L}_{\eta^8 \pi^+ \pi^- \gamma}$ and
${\cal L}_{\eta^1 \pi^+ \pi^- \gamma}$ are proportional to $N_c$, the vertex
${\cal L}_{\eta \pi^+ \pi^- \gamma}$ involving the physical $\eta$-meson is 
also proportional to $N_c$. Hence, the width of the decay 
$\eta \rightarrow \pi^+ \pi^- \gamma$ is proportional to $N_c^2$ and the 
observed width indeed implies $N_c = 3$. This decay should hence replace 
the textbook example $\pi^0 \rightarrow \gamma \gamma$ for lending experimental
support to the fact that there are three colors in our world.

Also the vertices
\begin{eqnarray}
\label{vertices}
&&{\cal L}_{\pi^0 K^0 \overline{K^0} \gamma} =
\frac{e (N_c - 1)}{8 \pi^2 F_\pi^3}
\varepsilon_{\mu\nu\rho\sigma} A_\mu \p_\nu \pi^0 \p_\rho K^0 \p_\sigma 
\overline{K^0}, \nonumber \\
&&{\cal L}_{\pi^0 K^+ K^- \gamma} = 
\frac{e (N_c + 1)}{8 \pi^2 F_\pi^3}
\varepsilon_{\mu\nu\rho\sigma} A_\mu \p_\nu \pi^0 \p_\rho K^+ \p_\sigma K^-,
\nonumber \\
&&{\cal L}_{\eta^8 K^0 \overline{K^0} \gamma} = 
\frac{e \sqrt{3} (1 - N_c)}{8 \pi^2 F_\pi^3}
\varepsilon_{\mu\nu\rho\sigma} A_\mu \p_\nu \eta^8 \p_\rho K^0 \p_\sigma 
\overline{K^0}, \nonumber \\
&&{\cal L}_{\eta^8 K^+ K^- \gamma} =
\frac{e \sqrt{3} (1 - N_c/3)}{8 \pi^2 F_\pi^3}
\varepsilon_{\mu\nu\rho\sigma} A_\mu \p_\nu \eta^8 \p_\rho K^+ \p_\sigma 
K^-, 
\end{eqnarray}
are explicitly $N_c$-dependent. However, for kinematic reasons these processes 
do not contribute to single particle decays and are hence more difficult to 
observe experimentally. In the microscopic theory the vertices of 
eq.(\ref{vertices}) result from quark box diagrams that are proportional to
\begin{eqnarray}
&&\mbox{Tr}(T^3 [U^+,U^-] Q) = - \frac{N_c}{4} Q_d = \frac{1}{8}(N_c - 1),
\nonumber \\
&&\mbox{Tr}(T^3 [V^+,V^-] Q) = \frac{N_c}{4} Q_u = \frac{1}{8}(N_c + 1),
\nonumber \\
&&\mbox{Tr}(T^8 [U^+,U^-] Q) = \frac{N_c}{4 \sqrt{3}} 3 Q_d = 
\frac{\sqrt{3}}{8}(1 - N_c), \nonumber \\
&&\mbox{Tr}(T^8 [V^+,V^-] Q) = \frac{N_c}{4 \sqrt{3}}(Q_u + 2 Q_d) = 
\frac{\sqrt{3}}{8}(1 - \frac{N_c}{3}),
\end{eqnarray}
respectively. Here we have used
\begin{equation}
V^\pm = \frac{1}{\sqrt{2}}(T^4 \pm i T^5), \
U^\pm = \frac{1}{\sqrt{2}}(T^6 \pm i T^7).
\end{equation}

\section{Summary and Conclusions}

We have considered a consistent standard model with arbitrary odd $N_c$ and 
with $N_u$ light up-type quarks as well as $N_d$ light down-type quarks. The 
partition function of the corresponding low-energy effective theory for the 
Goldstone bosons of the strong interactions in the background of an
electromagnetic gauge field then takes the form
\begin{eqnarray}
Z[A_\mu]&=&\int {\cal D}U \exp(- S[U,A_\mu]) \exp(i N_c S_{WZW}[U,A_\mu])
\nonumber \\
&\times&\exp(i (1 - N_c \frac{N_d - N_u}{N_f}) S_{GW}[U,A_\mu]).
\end{eqnarray}
In the two flavor case ($N_f = 2$), the Wess-Zumino-Witten term reduces to 
$\mbox{Sign}[U] = \pm 1 \in \Pi_4[SU(2)] = \Z(2)$ (for odd $N_c$) and thus 
becomes $N_c$-independent. With $N_u = N_d = 1$ the Goldstone-Wilczek term also
becomes $N_c$-independent, and hence $N_c$ does not appear explicitly in the 
low-energy effective theory of the standard model. In particular, the width of 
the decay $\pi^0 \rightarrow \gamma \gamma$, which is entirely due to the 
Goldstone-Wilczek term, is not proportional to $N_c^2$. Also other photon-pion 
vertices, like ${\cal L}_{\pi^0 \pi^+ \pi^- \gamma}$, do not depend on $N_c$ 
explicitly. Of course, the low-energy effective theory still depends implicitly
on the number of colors, because quantities like $F_\pi$ are $N_c$-dependent. 
Hence, if one computes $F_\pi$ in a nontrivial lattice QCD calculation with 
$N_c$ colors, and then compares, for example, with the observed 
$\pi^0 \rightarrow \gamma \gamma$ decay width, one will correctly conclude that
$N_c = 3$ in our world. However, if one takes the value of $F_\pi$ from 
experiment, it is impossible to literally see the number of colors by detecting
the photons emerging from the decay of the neutral pion.

In the three flavor case with $N_u = 1$, $N_d = 2$  ($N_u + N_d = N_f = 3$) and
$N_c = 3$ the Goldstone-Wilczek term vanishes and the contribution to the 
$\pi^0$ decay seems to be entirely due to the Wess-Zumino-Witten term. However,
for general $N_c$, the $N_c$-dependences of both terms cancel and the resulting
width for the decay $\pi^0 \rightarrow \gamma \gamma$ stems from the 
$N_c$-independent part of the Goldstone-Wilczek term only. The cancellations of
the $N_c$-dependent terms are not limited to the vertex 
${\cal L}_{\pi^0 \gamma \gamma}$, but appear for all electromagnetic processes
involving only pions and photons. Still, for $N_f \geq 3$ there are indeed some
processes that allow one to literally see the number of colors. For example, 
the width of the decay $\eta \rightarrow \pi^+ \pi^- \gamma$ is proportional to
$N_c^2$ and the observed width indeed implies that there are three colors in
our world. This decay should hence replace the textbook process 
$\pi^0 \rightarrow \gamma \gamma$ lending experimental support to $N_c = 3$.

It should be noted that our discussion does not apply to very large values of
$N_c$. In that case, the $\eta'$-meson becomes light and should be included in
the chiral Lagrangian \cite{DiV80,Wit80,Kai00}. It would be interesting to 
repeat our arguments in that situation, especially in order to further address
the issue of meson mixing in the decay $\eta \rightarrow \pi^+ \pi^- \gamma$. 
It might also be worthwhile to reconsider those calculations that hold the 
quark charges fixed while taking the large $N_c$ limit. When one uses the 
consistent quark charges of eq.(\ref{charges}), one would expect to obtain 
results for electroweak processes with a more well-behaved dependence on the 
number of colors.

As one would have expected, at the end of this paper we still conclude that in 
our world $N_c = 3$. However, we have sometimes been taught to believe this 
fact for the wrong reasons. We conclude this paper by expressing our hope that 
in the future some textbooks will reflect the results of the discussion 
presented here.

\section*{Acknowledgements}

We like to thank E. Farhi and J. Goldstone for illuminating discussions. 
U.-J. W. also thanks the students of the spring 2001 MIT course 8.325 for 
asking some of the questions that have led to this paper.

\end{document}